\documentclass[aps,prl,twocolumn,showpacs,groupedaddress,letterpaper,fleqn]{revtex4}
\usepackage{graphicx}% Include figure files
\usepackage{dcolumn}% Align table columns on decimal point
\usepackage{bm}% bold math

\newcommand{\pt}{\mbox{$p_T$}}                  %pT
\newcommand{\ptrel}{\mbox{$p_T^{\rm rel}$}}     %PtRel
\newcommand{\ppbar}{\mbox{$p\overline{p}$}}     %ppbar
\newcommand{\mupm}{\mbox{$\mu^+\mu^-$}}         %mu+mu-
\newcommand{\jpsi}{\mbox{$J/\psi$}}             %J/psi
\newcommand {\ra}        {\rightarrow}          % -->

\newcommand{\gevcc}{\mbox{GeV/$c^2$}}           %GeV/c^2
\newcommand{\gevc} {\mbox{GeV/$c$}}             %GeV/c
\newcommand{\mevcc}{\mbox{MeV/$c^2$}}           %MeV/c^2
\newcommand{\ifb}{\mbox{fb$^{-1}$}}   %inverse femtobarns

\begin{document}
%%%%%%%%%%%%%%%%%%%%%%%%%%%%%%%%%%%%%%%%%%%%%%%%%%%%%%%%%%%%%
% the following line is for submission
\hspace{5.2in} \mbox{Fermilab-Pub-08/047-E}
%%%%%%%%%%%%%%%%%%%%%%%%%%%%%%%%%%%%%%%%%%%%%%%%%%%%%%%%%%%%%

\title{
Measurement of the \bm{$B_c$} meson mass in the exclusive decay 
\bm{$B_c\ra\jpsi~\pi$}
}

% use the official authorlist for publication

% LIST_OF_AUTHORS_R2.TEX               2/19/08              
%
\author{V.M.~Abazov$^{36}$}
\author{B.~Abbott$^{75}$}
\author{M.~Abolins$^{65}$}
\author{B.S.~Acharya$^{29}$}
\author{M.~Adams$^{51}$}
\author{T.~Adams$^{49}$}
\author{E.~Aguilo$^{6}$}
\author{S.H.~Ahn$^{31}$}
\author{M.~Ahsan$^{59}$}
\author{G.D.~Alexeev$^{36}$}
% ...

\affiliation{\vspace{0.1 in}(The D\O\ Collaboration)\vspace{0.1 in}}
% ...
\affiliation{$^{6}$University of Alberta, Edmonton, Alberta, Canada,
                Simon Fraser University, Burnaby, British Columbia, Canada,
                York University, Toronto, Ontario, Canada, and
                McGill University, Montreal, Quebec, Canada}
% ...
\affiliation{$^{29}$Tata Institute of Fundamental Research, Mumbai, India}
% ...
\affiliation{$^{31}$Korea Detector Laboratory, Korea University, Seoul, Korea}
% ...
\affiliation{$^{36}$Joint Institute for Nuclear Research, Dubna, Russia}
% ...
\affiliation{$^{49}$Florida State University, Tallahassee, Florida 32306, USA}
% ...
\affiliation{$^{51}$University of Illinois at Chicago,
                Chicago, Illinois 60607, USA}
% ...
\affiliation{$^{59}$Kansas State University, Manhattan, Kansas 66506, USA}
% ...
\affiliation{$^{65}$Michigan State University,
                East Lansing, Michigan 48824, USA}
% ...
\affiliation{$^{75}$University of Oklahoma, Norman, Oklahoma 73019, USA}
% ...

\date{February 28, 2008}
           
\begin{abstract}
      A fully reconstructed $B_c\ra\jpsi+\pi$ signal is observed 
      with the D0 detector at the Fermilab Tevatron \ppbar\ collider.
      Using 1.3 \ifb\ of integrated luminosity, the signal is extracted 
      with a significance more than five standard deviations above 
      background. The measured $B_c$ meson mass is 
      $6300
      \pm 14 \thinspace \mathrm{(stat)}
      \pm 5  \thinspace \mathrm{(sys)}$ \mevcc.
\end{abstract}

% Valid PACS numbers (see {\tt http://www.aip.org/pacs/}) may be entered 
% using the \verb+\pacs{#1}+ command. They are required for PRL/PRD papers.
% activate the following line for publication
%\pacs{Valid PACS appear here} 
% decays hadr and lept :  decays, bottom and charm
\maketitle
%%\newpage
%%\section{\label{sec:intro}Introduction}

$B_c$ mesons are predicted by the quark model to be members
of the $J^P=0^-$ pseudo-scalar ground-state multiplet and to
have zero isospin as the lowest-lying bound state of a
bottom anti-quark and a charm quark \cite{quigg}. 
$B_c$ properties are of special 
interest because of this meson's unique status as a short-lifetime
bound state of heavy but (unlike quarkonia) different flavor quarks. 
Measurements of its mass, production, and decay therefore allow for tests
of theories under new approximation regimes or extended 
validity ranges beyond quarkonia.

This analysis uses data collected by the D0 detector between 
April 2002 and March 2006 at the Fermilab Tevatron \ppbar\ collider 
operating at $\sqrt s = 1.96$~TeV. The data sample corresponds to 
approximately 1.3 \ifb\ of integrated luminosity.  At the Tevatron the 
most easily identified decay modes of the $B_c$ have a \jpsi\ meson in 
the final state %%(with $\jpsi\ra\elpm$ for lepton triggers) 
and are either the semileptonic mode 
$B_c \ra \jpsi \ell \nu ~~(\ell = e, \mu)$, 
a signal with much higher statistics and thus more suitable for lifetime 
measurements, or the hadronic mode $B_c \ra \jpsi \pi$, 
more suitable for mass measurements given its fully exclusive 
reconstruction without the loss of an escaping neutrino.

The CDF collaboration has published results on both decay modes 
\cite{cdfmass,cdftau}, %%and has an updated $B_c$ mass measurement 
and has recently updated the $B_c$ mass measurement to
      $M(B_c) = 6275.6
      \pm 2.9 \thinspace \mathrm{(stat)}
      \pm 2.5 \thinspace \mathrm{(sys)}$ \mevcc\ \cite{cdfnew}.
%%The D0 collaboration has updated its $B_c$ lifetime measurement,
%%\cite{d0tau}, 
The present Letter is the first report by the D0 collaboration of a fully
reconstructed hadronic decay mode of this state.
The measured lifetime \cite{cdftau}
is consistent with the expectation of a shorter $B_c$ lifetime than for
other $B$ mesons due to the presence of a charm quark. The $B_c$ mass
has been predicted by various theoretical models \cite{thy}
and most recently \cite{lqcd}
with a three-flavor (unquenched) lattice QCD numerical algorithm
that yielded the smallest theoretical uncertainty, with the result 
$M(B_c) = 6304 \pm 12${\scriptsize $\begin{array}{c} + 18 \\ 
                                  - 0\end{array}$} \mevcc,
where the first error is the sum in quadrature of statistical and 
systematic uncertainties, and the second is due to heavy quark 
discretization effects.

The D0 detector is described elsewhere \cite{d0det}, 
and the elements most relevant to this analysis are the
tracking detectors inside a 2 T superconducting solenoidal magnet and the
muon detection chambers. 
For enhanced preselection efficiency, no specific trigger requirements
are applied, but all events satisfy one of a suite of muon triggers, 
typically requiring at least one %%(centrally matched) 
muon with %%of good track quality and 
transverse momentum (\pt) above 3 \gevc. The decay 
under study consists of a single detached secondary three-track vertex: 
$B_c\ra\jpsi \pi\ra\mupm\pi$ 
(charge conjugate modes, $\pi^\pm$, are always implied).
%%Events are reconstructed using the standard D0 software suite. 
Initial track selection extends to a pseudorapidity of $|\eta| < 2.0$
(where $\eta = -\ln[\tan(\theta/2)]$, and $\theta$ is the polar angle
with respect to the beam line), and rejects tracks with $\pt < 1.5$ \gevc.
Selected final state tracks must satisfy quality requirements based 
on established minimal hit patterns and a goodness of track fit.
Tracks identified as muons must have matching hits in all three layers 
of the muon detector.

Event selection starts with the requirement of an opposite-charge muon pair
that forms a common vertex and whose mass is consistent with that of the 
\jpsi\ meson (between 2.85 and 3.35 \gevcc). There follows a search for a 
third track that, together with the muons, must form a common vertex with
$\chi^2 < 16.0$ for the three degrees of freedom.
The \jpsi\ candidate must have $\pt > 4$ \gevc, and
the third particle is assigned the pion mass. Thus formed, the $B_c$
meson candidate is required to have %%a transverse momentum above 5 \gevc.
$\pt > 5$ \gevc.

Further $B_c$ candidate selection places constraints on quantities that 
proved to be strong discriminators against combinatoric backgrounds. 
The impact parameter (IP) significance of any particle, reconstructed 
either from a single track or a combination of tracks, is
$
I_{\text{sig}} = 
\sqrt{
[\epsilon_T/\sigma(\epsilon_T)]^2 + [\epsilon_L/\sigma(\epsilon_L)]^2
}
$,
where $\epsilon_T$ ($\epsilon_L$) is the transverse (longitudinal)
projection (with respect to the beam direction) of the track IP 
relative to the \ppbar\ interaction vertex, or primary vertex, 
and $\sigma$ is the associated uncertainty. The primary vertex is 
determined event by event using a method described in Ref. \cite{delphi}.
The transverse decay length significance of a decay (or secondary) vertex is
$
S_{xy} = L_{xy}/\sigma(L_{xy})
$
where $L_{xy}$ is the distance separating that vertex from the beam line.
The pointing cosine, $C_{xy}$, measures the alignment between $\vec L_{xy}$
and the transverse momentum direction of the decaying candidate particle. 
The isolation ${\cal{I}}$
of a $B_c$ candidate is defined as the ratio of two \pt\ sums: that from
the three candidate tracks, divided by that from all tracks with \pt\
above 0.3 \gevc\ whose momenta are lying within a cone of radius
$
\Delta{\cal{R}} = 
\sqrt{(\Delta\eta)^2 + (\Delta\phi)^2}
 = 0.5
$,
where $\Delta\eta$ and $\Delta\phi$ are distances in pseudorapidity and 
azimuthal angle from the $B_c$ momentum axis, respectively.

Throughout the background reduction process, a control procedure is
used that tests the effect of each discriminator against a well-understood 
signal sample, either reconstructed $B^\pm \ra \jpsi K^\pm$ candidates in 
data or candidates in a $B_c^\pm \ra \jpsi \pi^\pm$
simulated Monte Carlo sample. The latter is generated using 
{\sc EvtGen} \cite{evtgen} interfaced with {\sc pythia} \cite{pythia}, 
followed by full modeling of the detector response with {\sc geant} 
\cite{geant} and event reconstruction exactly as in data.

\jpsi\ candidates are mass constrained, i.e., their daughter muon 
momenta are corrected to yield the PDG \cite{pdg} mass value.
When the third track is assumed to be a kaon, a clean, high-statistics 
$B^\pm$ signal in invariant mass is observed in the data.  
This decay has a topology similar to the $B_c$ signal and is used as a 
reference in an initial round of selection cuts shown in Table 
\ref{tab:cuts} as Stage 1.
Here the $B^\pm$ signal and sideband regions are used as efficiency and 
rejection indicators of where to set selection thresholds. 
The $B^\pm$ study region extends from 4.98 to 5.58 \gevcc\ in invariant
mass, and the signal 
region is approximately $\pm 2\sigma$ wide from 5.20 to 5.36 \gevcc.
Individual cuts are required to be about 95\% efficient, with
typical background rejection of approximately 20\%.
The resulting thresholds are listed in Table \ref{tab:cuts}.

%%%%%%%%%%%%%%%%%%%%%%%%%%%%%%%%%%%%%%%
\begin{table}
\caption{Discriminators and their values at the two selection stages
(see text). $\ptrel(\pi)$ is introduced only for the second stage and
represents the transverse momentum of the pion candidate with respect
to the total $B_c$ candidate momentum.}
%%\begin{center}
\begin{ruledtabular}
\begin{tabular} {lccr}
Discriminator & Condition & ~Stage 1 & ~Stage 2 \\
\hline
$I_{\text{sig}}(B_c)$    & $<$ & 3.5 & 3.5  \\
$I_{\text{sig}}(\pi)$    & $>$ & 3.0 & 3.5  \\
\hline
$S_{xy}$           & $>$ & 3.0  & 4.5  \\
$C_{xy}$           & $>$ & 0.95 & 0.95 \\
${\cal{I}}$        & $>$ & 0.5  & 0.64 \\
\hline
\pt($\pi$)~~(\gevc)  & $>$ & 1.8 & 2.2 \\
$\ptrel(\pi)$ (\gevc)& $>$ & -- & 1.5 \\
$\ptrel(\pi)$ (\gevc) & $<$ & -- & 2.5 \\
\end{tabular}
\label{tab:cuts}
%%\end{center}
\end{ruledtabular}
\end{table}
%%%%%%%%%%%%%%%%%%%%%%%%%%%%%%%%%%%%%%%

However, there are differences between the $B^\pm$ and $B_c$.
Due to the lower (by about 1 \gevcc) invariant mass and the longer 
($b$-like versus $c$-like) lifetime of the $B^\pm$, background 
reduction undergoes a second stage, in which the $B_c$ Monte Carlo 
is used to model the signal. This second selection stage (Stage 2 in 
Table \ref{tab:cuts}) aims at re-optimizing, if needed, those cuts 
associated with $B_c$ specific decay properties. With the third track 
now assumed to be a pion, the range in invariant mass from 5.6 to 7.2 
\gevcc\ is
studied. A sub-range between 6.1 and 6.5 \gevcc\ is treated as the $B_c$ 
signal search window, and its invariant mass distribution in data is
kept blinded throughout the analysis. This sub-range is approximately 
$\pm 3\sigma$ (mass resolution as determined from simulation) wide, 
and covers both the theory expectations for the $B_c$ mass \cite{lqcd} 
as well as the observed values quoted in \cite{cdfmass,cdfnew}.
Data in mass sidebands outside this sub-range are used as a model 
for backgrounds and to quantify background rejection.
Table \ref{tab:cuts} lists those selections that were re-optimized
(or introduced, in the case of $\ptrel(\pi)$) in Stage 2, and summarizes 
their evolution between the two selection stages. At this stage there
remain no dimuon vertices with more than one candidate for the third 
track, and no events with more than one $B_c$ candidate.

%%%%%%%%%%%%%%%%%%%%%%%%%%%%%%%%%%%%%%%%%%%%%%%%%%%%%%%%%
\begin{figure}
\includegraphics[scale=0.42]{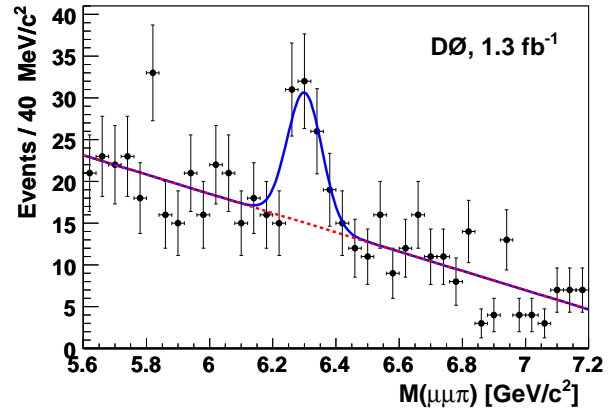}
\caption{\label{fig:plot}
$\jpsi \pi$ invariant mass distribution of $B_c$ candidates after the final 
selection.  A projection of the unbinned maximum likelihood fit to the 
distribution is shown overlaid.
}
\end{figure}
%%%%%%%%%%%%%%%%%%%%%%%%%%%%%%%%%%%%%%%%%%%%%%%%%%%%%%%%%

From $B_c$ simulated events, the $B_c$ mass signal is found to be 
well-modeled by a Gaussian function with a width of 55 \mevcc.  The mass 
resolution of the $B^\pm \ra \jpsi K^\pm$ signal observed in the data 
under similar conditions, after all selections have been applied,
reproduces the same width when scaled by the ratio of the $B^\pm$ 
and $B_c$ masses.

The resulting $\jpsi \pi$ invariant mass is shown 
in Fig.\ \ref{fig:plot} where a clear excess is seen near 6.3 \gevcc. 
An unbinned maximum log-likelihood (UML) fit of the $\jpsi \pi$ invariant
mass distribution is performed, where the signal is 
modeled by a Gaussian function with width fixed to a value of 55 \mevcc, 
and combinatoric backgrounds are modeled by a first-degree polynomial.
The result of the UML fit is overlaid in Fig.\ \ref{fig:plot} and yields 
a signal of $54 \pm 12$ events and a $B_c$ mass value of 
6300.7 $\pm$ 13.6 \mevcc. 
To estimate the signal significance, the same fit is repeated under the
assumption that no signal is present. From the negative log-likelihoods
of the signal plus background and background-only hypotheses, the
signal significance is extracted \cite{pdg} as
$
N_\sigma = \left\{ 2\ln[
           {{\cal L}(s+b)}/{{\cal L}(b)}]
           \right\}^{1/2}
         = 5.2
$
standard deviations above background. For another estimate of 
signal significance, $\chi^2$ fits to data (in the 40 \mevcc\ bins
of Fig.\ \ref{fig:plot}) under both hypotheses %%(s+b and b-only)
produce an increase in fit $\chi^2$ of 27 units, again indicating
$N_\sigma = 5.2$ standard deviations above background.

Possible biases and systematic uncertainties affecting the $B_c$ mass
determination are estimated using both the $B^\pm$ signal in the data
and the $B_c$ signal in either the data or the simulation.
Uncertainty assessments are made as these samples 
are re-fitted under various test hypotheses.
Sources of systematic uncertainties are the event selection,
the fitting procedure (input mass resolution and data modeling), and
the reconstructed mass scale.

The fitted mass values are examined in the simulated signal sample
as the value of the $\pt(\pi)$ 
threshold is varied from 1.9 to 2.5 \gevc.  No systematic mass bias is 
observed, but statistical fluctuations of $\pm 4.0$ \mevcc\ are observed 
and assigned as a systematic uncertainty.
Similarly, the \ptrel\ lower threshold is varied between no cut and 2.0
\gevc, and the resultant mass variation indicates a small upward mass 
bias of 0.5 \mevcc\ for the cut value adopted with respect to the no
cut case. The observed $B_c$ mass is corrected accordingly, and a 
100\% uncertainty is assigned to this correction. There is no
indication of a bias in mass due to the upper \ptrel\ limit.

The values of the selection cuts that are not directly 
related to the kinematics of the third particle (the pion or kaon 
candidates in the $B_c$ or $B^\pm$ cases, respectively) are varied 
within reasonable values.  No mass biases are observed,
and from the range of mass values obtained, a systematic uncertainty 
of $\pm 2.5$ \mevcc\ is assigned due to the choice of these selection cuts.

%%%%%%%%%%%%%%%%%%%%%%%%%%%%%%%%%%%%%%%
\begin{table}
\caption{\label{tab:unc}
Summary of systematic uncertainties in the $B_c$ mass measurement.
}
%%\begin{center}
\begin{ruledtabular}
\begin{tabular} {lccr}
Source & Component & Value (\mevcc) \\
\hline
Selection       & $\pi$ kinematics & 4.0 \\
                & other        & 2.5 \\
%%                & sub-total   & 5.0 \\
\hline
Data modeling   & mass resolution  & 0.6  \\
                & background model & 0.5  \\
                & signal shape     & 0.5  \\
%%                & sub-total        & 0.9  \\ 
\hline
Mass scale  &     & 1.0 \\
\hline
Total           &     & 4.9 \\
\end{tabular}
%%\end{center}
\end{ruledtabular}
\end{table}
%%%%%%%%%%%%%%%%%%%%%%%%%%%%%%%%%%%%%%%

To assess the systematic uncertainty due to the uncertainty of the mass 
resolution, the width of the Gaussian is allowed to float in the fit. 
The width input is also changed from the nominal value of 55 \mevcc\
to other fixed values in the range from 45 to 65 \mevcc.  From the 
variation of fitted mass results, a value of $\pm 0.6$ \mevcc\
is assigned to this uncertainty.

The background model is changed from a first-degree 
polynomial to a second-degree and third-degree polynomial, and to an 
exponential function.  From the resulting change in mass observed, a 
systematic uncertainty of $\pm 0.5$ \mevcc\ is assigned due to uncertainty 
in the background model.
The signal model is changed from a single Gaussian to a double 
Gaussian function, and the resulting shift of 0.5 \mevcc\ is assigned as a 
systematic uncertainty. 

Lastly, for an estimate of the mass scale uncertainty, a direct 
comparison is carried out between generated
and reconstructed Monte Carlo masses, as well as between recent D0 mass 
measurements of well-known $B$ states and the world averages of their 
measurements \cite{pdg}. From the observed range of mass differences, 
a systematic uncertainty of $\pm 1.0$ \mevcc\ is assigned due to uncertainty 
in the D0 mass scale for the $B_c$ decay.

A summary of all systematic uncertainties
in the $B_c$ mass measurement is shown in Table \ref{tab:unc}.
The overall systematic uncertainty is $\pm 4.9$ \mevcc. The mass fit
result of 6300.7 $\pm$ 13.6 \mevcc\ is corrected by $-0.5$ \mevcc\ for the
\ptrel\ bias. The final result for the $B_c$ mass is 
      $6300
      \pm 14 \thinspace \mathrm{(stat)}
      \pm 5  \thinspace \mathrm{(sys)}$ \mevcc.

In summary, using a dataset corresponding to 1.3 \ifb, a signal for 
$B_c\ra\jpsi \pi$ has been observed with a significance higher than
five standard deviations above background.  The mass of the $B_c$ meson 
has been measured and found to be consistent with the latest and
most precise lattice QCD prediction \cite{lqcd}.
Besides its relevance in the development and tuning of heavy-quark
bound-state models, the $B_c$ sample described here, with added
integrated luminosity, is expected to be used in the extraction of
lifetime, relative branching ratio, and production rate.
\begin{acknowledgments}
% acknowledgement_paragraph_r2.tex                         2/19/08
%
We thank the staffs at Fermilab and collaborating institutions, 
and acknowledge support from the 
DOE and NSF (USA);
CEA and CNRS/IN2P3 (France);
FASI, Rosatom and RFBR (Russia);
CNPq, FAPERJ, FAPESP and FUNDUNESP (Brazil);
DAE and DST (India);
Colciencias (Colombia);
CONACyT (Mexico);
KRF and KOSEF (Korea);
CONICET and UBACyT (Argentina);
FOM (The Netherlands);
STFC (United Kingdom);
MSMT and GACR (Czech Republic);
CRC Program, CFI, NSERC and WestGrid Project (Canada);
BMBF and DFG (Germany);
SFI (Ireland);
The Swedish Research Council (Sweden);
CAS and CNSF (China);
and the
Alexander von Humboldt Foundation.

\end{acknowledgments}

\end{document}
%
% ****** End of file apssamp.